\documentclass[aps,twocolumn,floatfix,superscriptaddress,showpacs]{revtex4-1}

\usepackage{graphicx}
\usepackage{amsmath}
\usepackage{amssymb}
\usepackage{hyperref}
\usepackage{soul}

\usepackage[pdftex]{color}

\newcommand{\ket}[1]{| #1 \rangle}
\newcommand{\bra}[1]{\langle #1 |}

\newcommand{\figref}[1]{Fig.~\ref{#1}}

\usepackage{float}

\begin{document}

\title{One-particle-density-matrix occupation spectrum of many-body localized states after a global quench}

\author{Tal{\'i}a L.~M. Lezama}
\affiliation{Max-Planck-Institut f\"ur Physik komplexer Systeme, 01187 Dresden, Germany}
\author{Soumya Bera}
\affiliation{Max-Planck-Institut f\"ur Physik komplexer Systeme, 01187 Dresden, Germany}
\affiliation{Department of Physics, Indian Institute of Technology Bombay, Mumbai 400076, India}
\author{Henning Schomerus}
\affiliation{Department of Physics, Lancaster University, LA1 4YB Lancaster, United Kingdom}
\author{Fabian Heidrich-Meisner}
\affiliation{Department of Physics and Arnold Sommerfeld Center for Theoretical Physics,
Ludwig-Maximilians-Universit\"at M\"unchen, 80333 M\"unchen, Germany}
\author{Jens H.\ Bardarson}
\affiliation{Max-Planck-Institut f\"ur Physik komplexer Systeme, 01187 Dresden, Germany}
\affiliation{Department of Physics, KTH Royal Institute of Technology, Stockholm, SE-106 91 Sweden}

\begin{abstract}
The emergent integrability of the many-body localized phase is naturally understood in terms of localized quasiparticles. 
As a result, the occupations of the one-particle density matrix in eigenstates show a Fermi-liquid-like discontinuity. 
Here we show that in the steady state reached at long times after a global quench from a perfect density-wave state, this occupation discontinuity is absent, reminiscent of a Fermi liquid at a finite temperature, while the full occupation function remains strongly nonthermal.
We discuss how one can understand this as a consequence of the local structure of the density-wave state and the resulting partial occupation of quasiparticles.
This partial occupation can be controlled by tuning the initial state and can be described by an effective temperature. 

\end{abstract}

\maketitle

{\it Introduction.}---One of the basic notions of condensed matter physics is adiabatic continuity~\cite{Anderson:1984td}.
The Fermi liquid, a major example, is adiabatically connected to the Fermi gas by slowly turning on interactions:
the ground state of the Fermi gas evolves into that of the Fermi liquid and low-energy excited states evolve into excited quasiparticle states with identical quantum numbers~\cite{Nozieres:1999wr,Coleman:2015vz}.
The quasiparticle density operators represent conserved quantities;
the Fermi-liquid Hamiltonian is diagonal in the quasiparticle basis but is not quadratic as it contains quasiparticle density-density interactions that represent the back-action of all the other particles on excitations. 
The Fermi liquid fundamentally relies on reduced scattering due to limited phase space, resulting from the Fermi-sphere structure of the ground state, and is therefore only a valid description at low temperatures compared with the Fermi temperature~\cite{Nozieres:1999wr,Coleman:2015vz}.

Closed interacting disordered quantum systems can exhibit many-body localization (MBL)~\cite{Basko:2006hh,Gornyi:2005fv}, resulting in an ideal insulator with vanishing conductivities at finite, or even infinite~\cite{Oganesyan:2007ex,Pal:2010gr}, temperatures.
This MBL insulator is adiabatically connected to the Anderson insulator and therefore shares many features with the Fermi liquid~\cite{Bera:2015jh,Bera:2016ti}.
In contrast to the Fermi liquid, where only the ground state and the lowly excited states are adiabatically connected to the free Fermi gas, in an MBL insulator {\it every} eigenstate is adiabatically connected to some eigenstate of the Anderson insulator---for example in  
fully many-body-localized systems~\cite{Huse:2014co} (with exceptions~\cite{Parameswaran:2106}) where the relation can be provided by a finite-depth quantum circuit~\cite{Bauer:2013jw}.
The MBL phase is thus an emergent integrable phase~\cite{Serbyn:2013cl,Huse:2014co,Vasseur:2016kh} characterized by conserved quasiparticle densities, which are the density operators of Anderson orbitals locally dressed by particle-hole excitations~\cite{Ros:2015ib,Imbrie:2016cs,Imbrie:2016ib,Rademaker:2016jf}.
The eigenstates are product states of these quasiparticles and therefore satisfy an area law of entanglement~\cite{Bauer:2013jw,Kjall:2014bd,Luitz:2015iv},  
necessarily violating the eigenstate thermalization hypothesis~\cite{Deutsch:1991ju,Srednicki:1994dl,Rigol:2008bf}.
The construction of the conserved quantities (commonly referred to as $l$-bits) is extensively studied~\cite{Chandran:2015cw,Kim:2014wl,Rademaker:2016jf,Monthus:2016im,He:2016wy,Inglis:2016gn,OBrien:2016he,Pekker:2017cx,Pekker:2016vn,Ros:2015ib,Imbrie:2016cs,Imbrie:2016ib}.
As in the Fermi liquid, the MBL Hamiltonian is diagonal in the quasiparticle basis, but contains quasiparticle density-density interaction 
terms, which are absent in the Anderson insulator~\cite{Serbyn:2013cl,Huse:2014co}.
These interaction terms give rise to dephasing in dynamics that results in a logarithmic growth of entanglement entropy~\cite{Znidaric:2008cr,Bardarson:2012gc,Serbyn:2013he} (for examples of other quantum information measures, see~\cite{Goold:2015jh,Devakul:2015hw,Geraedts:2016fh,Singh:2016iza,Bera:2016gu,Iemini:2016cp,DeTomasi:2017fx,Campbell:2016uh}), and a slow relaxation of observables towards nonergodic stationary states at long times~\cite{Serbyn:2014dl}.
The adiabatic connectivity of the MBL phase to the Anderson insulator relies on the stability of Anderson localization against interactions~\cite{Basko:2006hh,Gornyi:2005fv,Ros:2015ib,Imbrie:2016cs,Imbrie:2016ib}.

This formal analogy between MBL and Fermi liquids was further developed in Refs.~\onlinecite{Bera:2015jh,Bera:2016ti}, which evinced a Fermi-liquid-like discontinuity in the eigenvalues of the one-particle density matrix (OPDM) in many-body eigenstates, analogous to a finite quasiparticle weight (see also Ref.~\onlinecite{Rademaker:2016tp}).
The discontinuity signals Fock-space localization, 
%
while the eigenvectors of the OPDM give localized orbitals, the natural orbitals, that can be used to construct an optimized single-particle approximation to the quasiparticles~\cite{Bera:2016ti}. 

In a Fermi liquid the occupation spectrum is discontinuous only at zero temperature; any nonzero temperature leads to a smooth occupation spectrum.
With the MBL eigenstates providing an analog to a zero-temperature Fermi liquid, it is natural to ask if there is also a finite-temperature analog.
We limit our consideration to temperature effects on quasiparticle occupations and assume that quasiparticle lifetimes are not affected. 
In this phenomenological analogy, in which each MBL eigenstate is a zero-temperature reference state, this requires partial occupations of quasiparticles compared with the reference occupations in a \emph{given} eigenstate. 
A generic combination of eigenstates, described by a mixed density matrix, does not work as this corresponds to summing over different random occupations of quasiparticles, or to mixing reference states.
Instead, we propose that a global quench from a local product state provides the physics we are after. 
Intuitively, a local density has a large overlap with some quasiparticle density.
An expansion of such a local density in the quasiparticles will therefore mainly contain the quasiparticles localized close-by, as if they were 
excited by a relatively small effective temperature. 
Initially we focus on a perfect density-wave state as the initial state.
Such a product state still has systematic phase differences between different quasiparticles unlike in thermal states.  
During time evolution, however, this quasiparticle superposition dephases such that the initial phase relationship is scrambled in the infinite-time steady state. 
The main result of our work is a characterization of this steady state with one-particle density matrix occupations that indeed mimic occupation effects of temperature in a Fermi liquid.  
We further show how the partial occupations in the steady state can be controlled by the structure of the initial state.

An initial density-wave state is also used in the ultracold atoms experiments that observed a finite imbalance between the density on even and odd sites as a signature of the absence of thermalization~\cite{Schreiber:2015jt,Bordia:2016dl,Bordia:2017bb} (see~\cite{Smith:2016cd,Choi:2016ic,Luschen:2017fz,Luschen:2016tz} for further experiments).
As a corollary result we therefore obtain a relation between the OPDM occupations and experiments.
In particular, we introduce an OPDM occupation imbalance, 
which behaves similar to the density imbalance but with a slower relaxation towards the steady state, thereby capturing dephasing.

{\it Model and methods.}---We study a system of spinless fermions hopping and repulsively interacting with their nearest neighbours in a disordered 1D lattice, with Hamiltonian
\begin{align}
H=J\sum_{i=1}^{L}\Big[ &-\frac{1}{2}(c_{i+1}^\dagger c_i + c_{i}^{\dagger}c_{i+1}) + \epsilon_{i}\Big(n_{i} - \frac{1}{2}\Big) \nonumber
\\ &+V\Big(n_{i} - \frac{1}{2}\Big)\Big(n_{i+1} - \frac{1}{2}\big)\Big],
\label{ham}
\end{align}
where $c_{i}^{\dagger}$ creates a fermion on site $i$ (among $L$ sites) and $n_{i} =  c_{i}^{\dagger}c_{i}$ is the number operator. 
Energies are expressed in terms of the hopping constant~$J$, whereas disorder and interaction strengths are denoted by the dimensionless quantities $W$ and $V$, respectively. 
The disorder is diagonal and taken from a box distribution $\epsilon_{i}\in [-W,W]$. 
We set $J=V=1$ throughout this work, in which case the localization-delocalization transition is found to be at $W_{c}=3.5\pm 1$ for energies in the middle of the spectrum~\cite{Pal:2010gr,Lev:2015co,Luitz:2015iv,Bera:2015jh,DeLuca:2013ba}.

Using exact diagonalization, we study the system described in \eqref{ham} for different system sizes $L$ and average over $10^4$ ($L=8,10,12$), $5\times10^3$ ($L=14$) and $4\times10^3$ ($L=16,18$) disorder realizations. 
We use periodic boundary conditions and fix the number of particles to half filling $N=L/2$. 
The symbol $\langle \cdot \rangle$ denotes the disorder average.

The initial state, unless stated otherwise, is a perfect density-wave state,
\begin{equation}
\ket{\Psi_{0}} = \prod_{i=1}^{L/2}c_{2i}^{\dagger}|0\rangle,
\end{equation}
which then evolves under the Hamiltonian~\eqref{ham} according to (we set $\hbar=1$)
$
|\Psi(t)\rangle = \exp(-iHt)\ket{\Psi_{0}}.
$
To characterize the state $|\Psi(t)\rangle$, we calculate the instantaneous OPDM
\begin{equation}
\rho_{ij}(t)=\langle \Psi(t) | c_{i}^{\dagger}c_{j} | \Psi(t) \rangle
\label{opdm}
\end{equation}
and diagonalize it. 
The eigenvalues $\{n_{\alpha}(t)\}$, with~$\alpha=1, 2, \dots, L$, are the occupations and the eigenfunctions $\{ | \phi_{\alpha} \rangle\}$ are the natural orbitals.
For each time, we order the OPDM eigenvalues in descending order $n_{1}(t)\geq n_{2}(t) \geq \dots \geq n_{L}(t)$, noting that the total particle number is conserved $\sum_{\alpha=1}^{L}n_{\alpha}(t) = \mathrm{tr}\, \rho(t) = N$ at all times.

\begin{figure}[tb]
\centering
\includegraphics[width=0.48\textwidth]{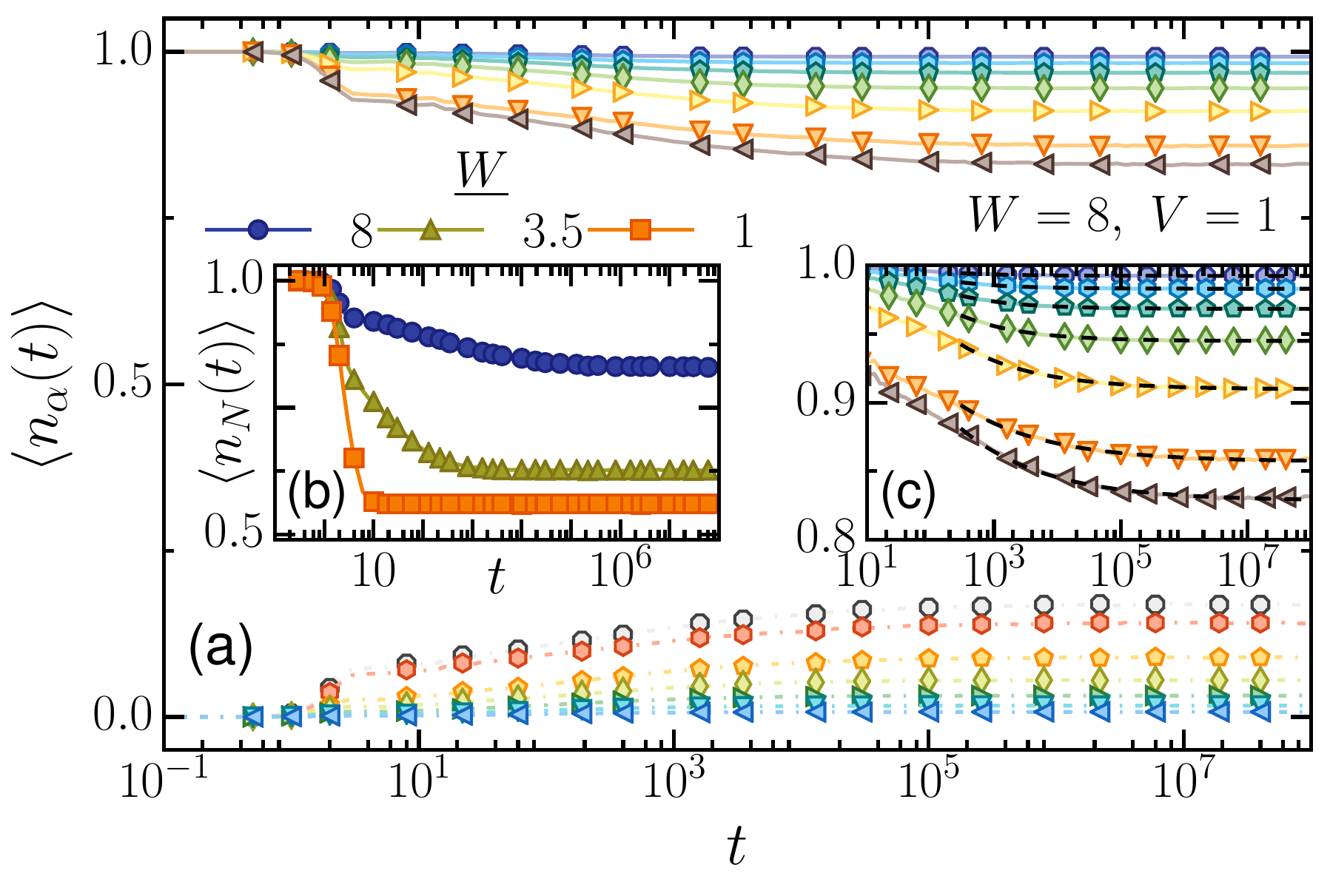}
\caption{(a) Evolution of the disorder-averaged occupation spectrum $\langle n_{\alpha}(t)\rangle$ deep in the MBL phase ($W=8$). (b) $\langle n_{N}(t)\rangle$, for both phases ($W=1,8$) and at the transition (W=3.5). (c)  Power-law relaxation (dashed lines, fits to the data) for the upper half of the spectrum. $L=14$ in (a-c).
}
\label{nat}
\end{figure}
{\it Evolution of occupations.}---We first address the nature of the relaxation dynamics of the occupations $\{n_{\alpha}(t)\}$.  
In the initial state $\ket{\Psi_{0}}$, half of the occupations are equal to one and the other half equal to zero, i.e.,  $n_{\alpha}(0)=1$ for $ \alpha\leq N$ and $n_{\alpha}(0)=0$ for $ \alpha\geq N+1$. 

\begin{figure*}[tb!]
\centering
\includegraphics[width=0.32\textwidth]{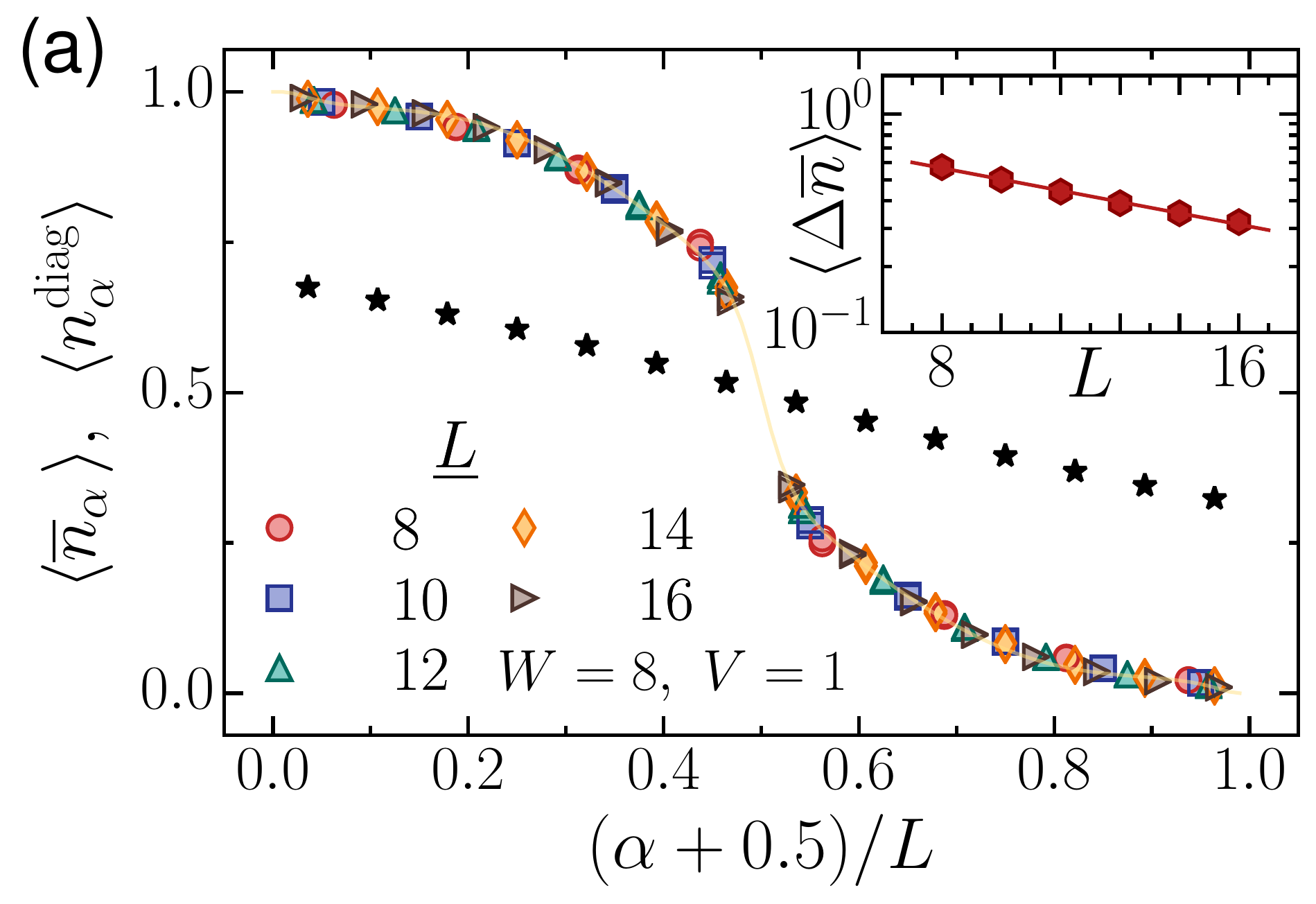}
\includegraphics[width=0.32\textwidth]{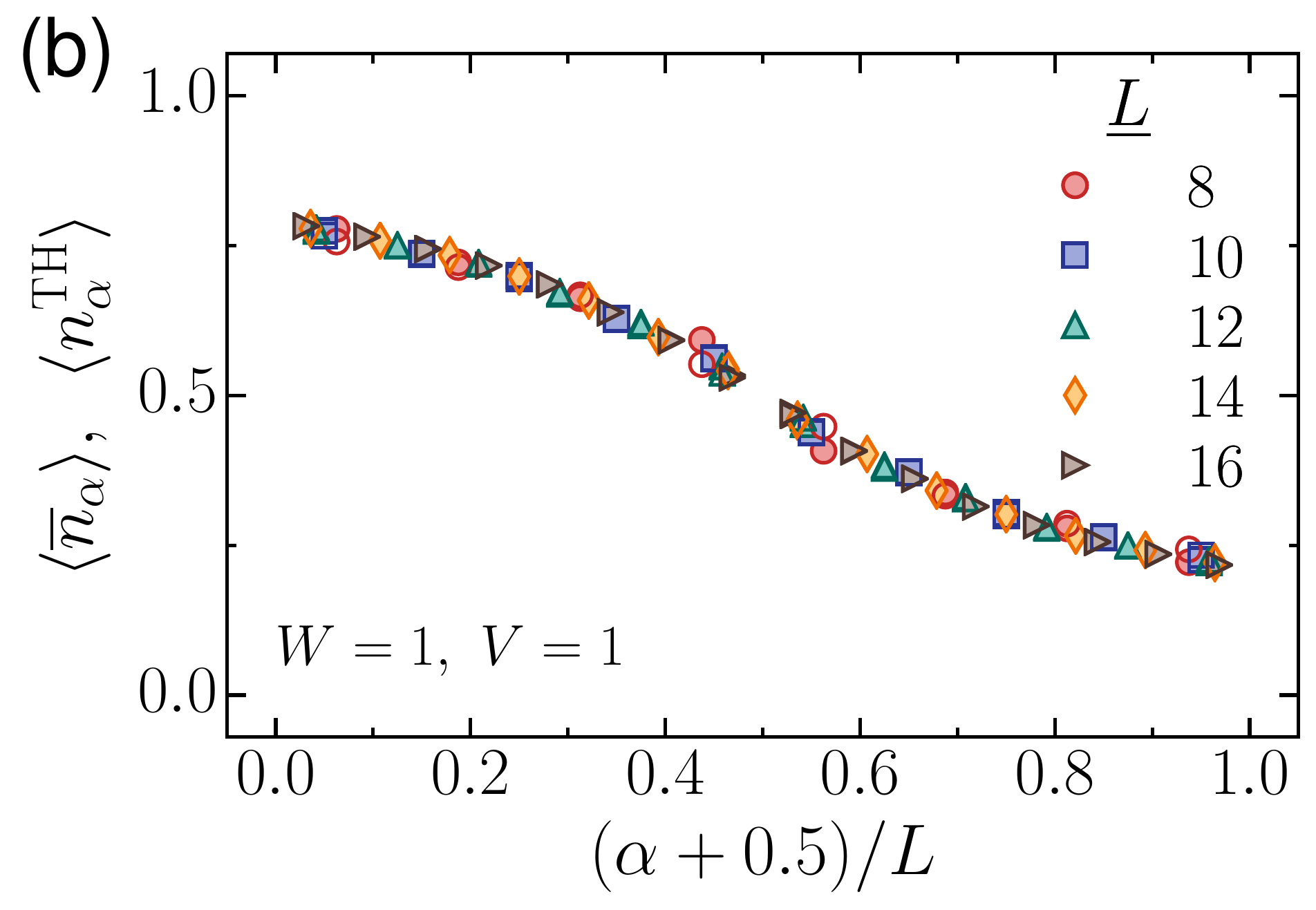}
\includegraphics[width=0.32\textwidth]{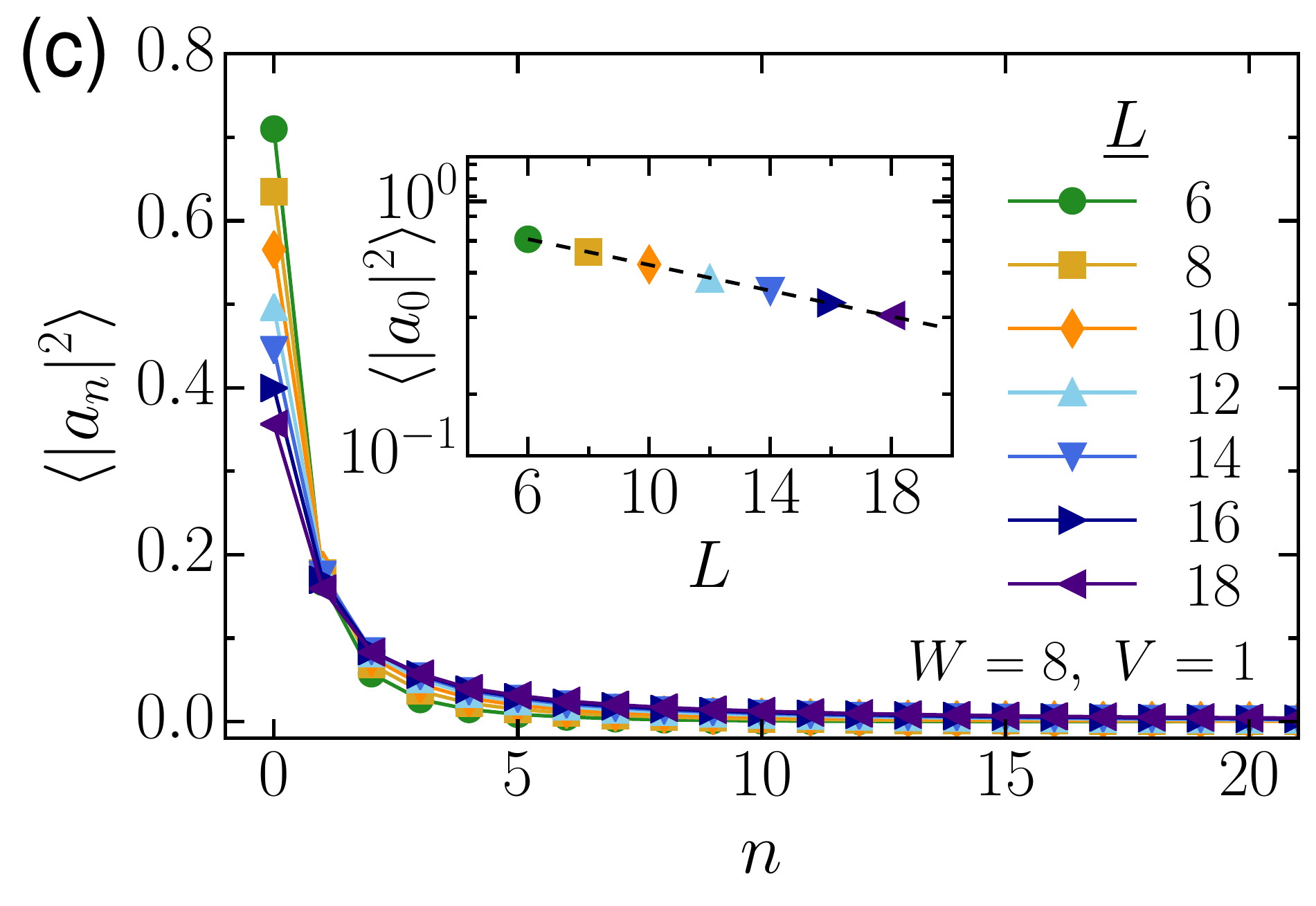}
\caption{Infinite-time and disorder-averaged distribution of occupations $\langle \overline{n}_{\alpha}\rangle$ as a function of system size $L$, for disorder strengths: (a) $W=8$, (b) $W=1$. The horizontal axis is scaled to $(\alpha+0.5)/L$. Additionally, (a) shows the diagonal-ensemble distribution $\langle {n}^{\mathrm{diag}}_{\alpha}\rangle$ for all $L$ (open symbols), the thermal ensemble $\langle n_{\alpha}^{\rm TH}\rangle$ for $L=14$ (stars) and the inset the discontinuity $\langle\Delta\overline{n}\rangle$ as a function of $L$ and a fit to an exponential $\sim e^{-0.06L}$.  Open symbols in (b) give the thermal distribution $\langle n_{\alpha}^{\rm TH}\rangle$ for all $L$. (c) shows the disorder-averaged overlap $\langle |a_{n}|^{2} \rangle$ between the initial state $\ket{\Psi_{0}}$ and the many-body eigenstates $\ket{n}$, plotted in decreasing order $|a_0|\geq|a_1|\geq\dots$ as a function of $n$ for $W=8$ and different $L$. Inset, disorder-averaged largest weight $\langle |a_{0}|^{2} \rangle$ as a function of $L$ as well as an exponential fit to exp$(-bL)$, with $b=0.06$.
}
\label{nai}
\end{figure*}

The time evolution of the occupation spectrum in the MBL phase is plotted in the main panel of~\figref{nat}.
Initially, the spectrum captures a fast expansion up to the localization length, followed by a slow relaxation in which the occupations approach their saturation values as a power law $\nu\, t^{-\gamma} + \delta$, starting at times of the order of $t \sim10^{2}$ (see \figref{nat}(c)).
The parameters $\delta$, $\nu$, and~$\gamma$ depend non-universally on $\alpha$ with the exponent $\gamma$ ranging between $0.3$ and $0.6$. 
In \figref{nat}(b), the time evolution of the occupation $\langle n_{N}(t) \rangle$ is shown for both phases and at the transition ($W\approx W_{c}$).
In the MBL phase ($W>W_{c}$), it undergoes a slow relaxation towards a nonthermal stationary state at long times ($t\sim 10^{8}$).
This slow relaxation is due to dephasing and is characteristic of the MBL phase~\cite{Serbyn:2014dl}. 
The instantaneous natural orbitals evolve from the initial onsite densities towards localized orbitals at long times, and the instantaneous occupations $\{n_{\alpha}(t)\}$ can therefore be seen as expectation values of local observables. 
In this sense, their approach to their stationary values is consistent with general arguments for  power-law relaxation of local observables \cite{Serbyn:2014dl}. 
In the ergodic phase ($W<W_{c}$), in contrast, we observe a fast relaxation towards a thermal stationary state. 

{\it Steady-state properties.}---On the basis of the above, it is natural to ask about the behavior of the occupation spectrum in the steady-state limit.
To this end, we explore the asymptotic behavior ($t\rightarrow\infty$) of the time-averaged density matrix, which we compare to density matrices that capture the separate effects of dephasing and thermalization. 
The steady-state density matrix at long times is described by the diagonal ensemble in both the MBL and the ergodic phase; only the latter is additionally reproduced by a thermal ensemble. 
Specifically, expanding $\ket{\Psi_0}=\sum_n a_n \ket{n}$ in terms of the many-body eigenstates $H\ket{n} = E_n\ket{n}$,  the time-evolved state takes the form
$
\ket{\Psi(t)} = \sum_n e^{-iE_nt}a_n\ket{n}
$,
and the density matrix is 
\begin{equation}
	\rho_{ij}(t) = \sum_{n,m} e^{-i(E_n-E_m)t}a_m^*a_n \bra{m}c_i^\dagger c_j\ket{n}.
\end{equation}
Taking the time average (denoted by $\bar{\cdot}$) yields
\begin{align}
\overline{\rho} &= \lim_{T\rightarrow\infty}\frac{1}{T-t_{0}}\int_{t_{0}}^{T} \rho(t) \, dt, \label{opdmta1}
\\
\bar{\rho}_{ij}&=  \sum_{n,m} \overline{e^{-i(E_n-E_m)t}}a_m^*a_n \bra{m}c_i^\dagger c_j\ket{n}.
\label{opdmta}
\end{align}
We take $t_0 = 10^5$ at which point the time evolution has reached a steady state.
For a nondegenerate system, the phases in~\eqref{opdmta} are random and sum to zero if $n\neq m$; therefore
\begin{equation}
\overline{\rho}_{ij} \approx \sum_n |a_n|^2 \bra{n}c^\dagger_ic_j\ket{n}\equiv \rho^{\mathrm{diag}}_{ij}.
\label{opdmd}
\end{equation}

It is important to contrast the time-averaged density matrix with the instantaneous occupations in~\figref{nat}.
The ordering of eigenvalues does not generally commute with time averaging, in particular if there is an interchange of occupations in the time evolution.
This can be expected to occur in the MBL phase, where the eigenvalues correspond to local quantities separated in space and consequently do not couple.
From now on, we therefore first time average the density matrix as in~\eqref{opdmta1} and only then determine and order its eigenvalues, denoted by $\bar{n}_{\alpha}$, in descending order.
The occupations $\langle\bar{n}_\alpha\rangle$ are plotted in~\figref{nai} as a function of $L$ and for three different values of $W$. In particular, we compare $\langle\bar{n}_\alpha\rangle$ with the ordered eigenvalues obtained directly from the diagonal ensemble~\eqref{opdmd}, denoted by $\langle{n}_{\alpha}^{\mathrm{diag}}\rangle$; both are plotted in~\figref{nai}(a) as a function of $L$ for $W=8$, with excellent agreement. 
In the ergodic phase, we further find good agreement with the eigenvalues of the thermal OPDM, $\langle{n}_{\alpha}^{\mathrm{TH}}\rangle$, obtained from $\rho_{\text{TH}} = \mathrm{tr}\left(\rho_{c} c_{i}^{\dagger}c_{j}\right)$,
where we use the density matrix of the canonical ensemble 
$\rho_{c} = e^{-\beta H}/\mathrm{tr}\,(e^{-\beta H})$, with inverse temperature $\beta$ set by the requirement that the energy of the state be 
$E = \langle\Psi_{0}|H|\Psi_{0} \rangle = \mathrm{tr} \left( \rho_c H \right)$ (see~\figref{nai}(b)).
The occupations obtained in the MBL phase are, in contrast, highly nonthermal as revealed by the comparison with the thermal occupation spectrum, plotted as stars in~\figref{nai}(a).
The OPDM occupations tend to exhaust the full range of  values between 0 and 1, similar to the occupations in eigenstates~\cite{Bera:2015jh,Bera:2016ti}, but with a discontinuity that vanishes exponentially as $L\to\infty$, see the inset in~\figref{nai}(a). 
This main result of our work suggests that a global quantum quench from a product state of local densities results in partial quasiparticle occupations and thus a continuous occupation spectrum, similar to the effect of a finite temperature in a Fermi liquid.

The absence of the discontinuity is best understood in the diagonal ensemble.
With the initial state being a product state of single-site occupations, it has a large overlap with the eigenstates that have a large weight on these sites.
The quenched state will, to first order in perturbation theory, inherit the step function from this eigenstate, while all the other states provide smearing of the step. 
To support this argument, we order the many-body eigenstates $\ket{n}$ according to their overlap with the initial state such that $|a_0| \geq  |a_1| \geq \cdots \geq |a_n|$, and in~\figref{nai}(c), we plot the disorder-averaged overlap $\langle |a_n|^{2} \rangle$ as a function of $n$.
This function decays quickly with $n$ and the largest overlap also decays exponentially with $L$. 
This is consistent with eigenstates built from $N$ quasiparticles each with an overlap with absolute value $c < 1$ with a given site density, and therefore, a total overlap that scales like $ |a_0|  \propto c^N$. 
The maximum-overlap eigenstate $\ket{0}$ has an OPDM $\rho^{(0)}$ with a zero-temperature Fermi-liquid-like step function.
The unitary transformation that diagonalizes $\rho^{(0)}$ approximately diagonalizes the OPDM $\rho^{(n)}$ of the higher eigenstates $\ket{n}$, but with random ordering such that the disorder average $\langle n_\alpha^{(n)} \rangle$ becomes a smooth function without any discontinuity (see Appendix A) for detailed calculations supporting this picture).
The resulting prediction of the diagonal ensemble for the discontinuity is then
\begin{equation}
\langle \Delta {n}^{\mathrm{diag}} \rangle \equiv  \langle {n}^{\mathrm{diag}}_{N} \rangle- \langle {n}^{\mathrm{diag}}_{N+1} \rangle  \approx \langle |a_0|^2\Delta n^{(0)}\rangle.
\label{deltad}
\end{equation}
This indeed goes to zero exponentially with $L$ since $\langle |a_0|^2 \rangle \sim e^{-bL}$, consistent with the inset to~\figref{nai}(a).

\begin{figure}[tb!]
\centering
\includegraphics[width=0.48\textwidth]{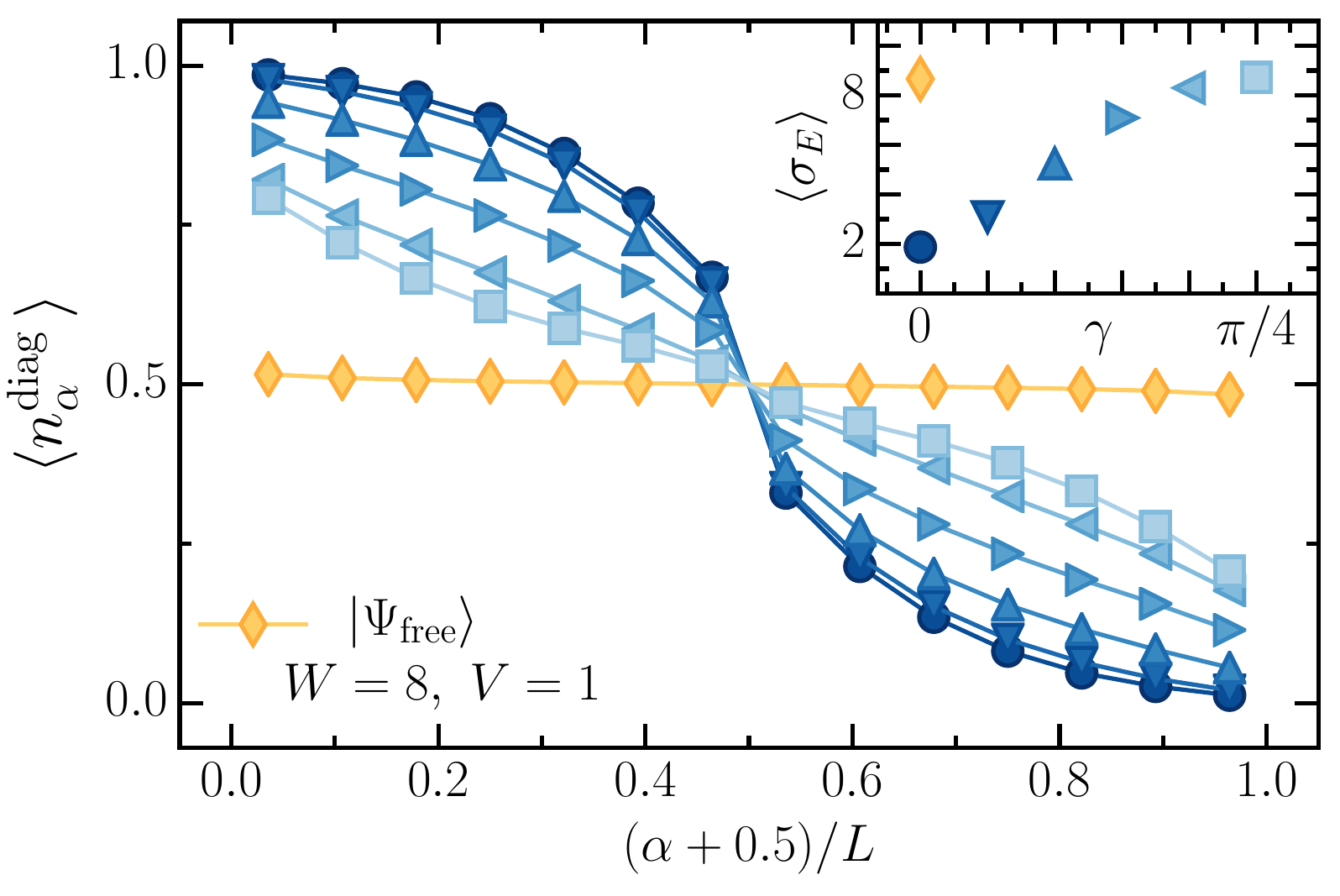}
\caption{Disorder-averaged occupations $\langle n_{\alpha}^{\mathrm{diag}}\rangle$ in the diagonal ensemble for different initial product states $|\Psi_{\gamma}\rangle$ corresponding to $\gamma= m\pi/20$ with $m=0,1,\dots,5$, as well as for the clean initial state $|\Psi_{\mathrm{free}}\rangle$ (diamonds), for values $W=8$ and $L=14$. Inset: the disorder-averaged standard deviation $\sigma_{E}$ of the energy in the initial state as a function of~$\gamma$ and in $|\Psi_{\mathrm{free}}\rangle$ (diamond). 
}
\label{temperature}
\end{figure}

{\it Tuning quasiparticle occupations.}---We can systematically tune the distance to a reference MBL eigenstate by generalizing the density-wave inital state to 
\begin{equation}
	\ket{\Psi_\gamma} = \prod_{i=1}^{L/2} [\cos(\gamma) c_{2i}^\dagger + \sin(\gamma) c_{2i+1}^\dagger]\ket{0}.
\end{equation}
In Fig.~\ref{temperature} we plot the steady-state occupation spectrum for several values of $\gamma$.
With increasing $\gamma$ the energy variance of the state $\ket{\Psi_\gamma}$ increases, corresponding to exciting quasiparticles and an increased effective temperature, which is reflected in the long-time occupation spectrum deviating more and more from the step function. 
This provides a systematic, and experimentally feasible~\cite{Folling:2007}, procedure to tune the quasiparticle occupations. 

While the energy variance provides a proxy for the effective temperature, by being a measure of the amount of quasiparticle excitations, it is important that these excitations are with respect to a definite reference eigenstate that acts as the ground state. 
This is guaranteed by the local structure of the initial state $\ket{\Psi_\gamma}$.
If we instead take an initial state without such a reference, for example a highly excited eigenstate $\ket{\Psi_{\mathrm{free}}}$ of the clean Hamiltonian \eqref{ham} with $W=0$, we obtain a practically flat distribution $\langle n_{\alpha}^{\mathrm{diag}} \rangle \approx 1/2 \,\, \forall \alpha$), see diamonds in \figref{temperature}.
This is the case even though the energy variance of this state is similar to that of $\Psi_\gamma$ with $\gamma = \pi/4$, and results from the mixing of reference states (see Appendix B). 

\begin{figure}[tb!]
\centering
\includegraphics[width=0.48\textwidth]{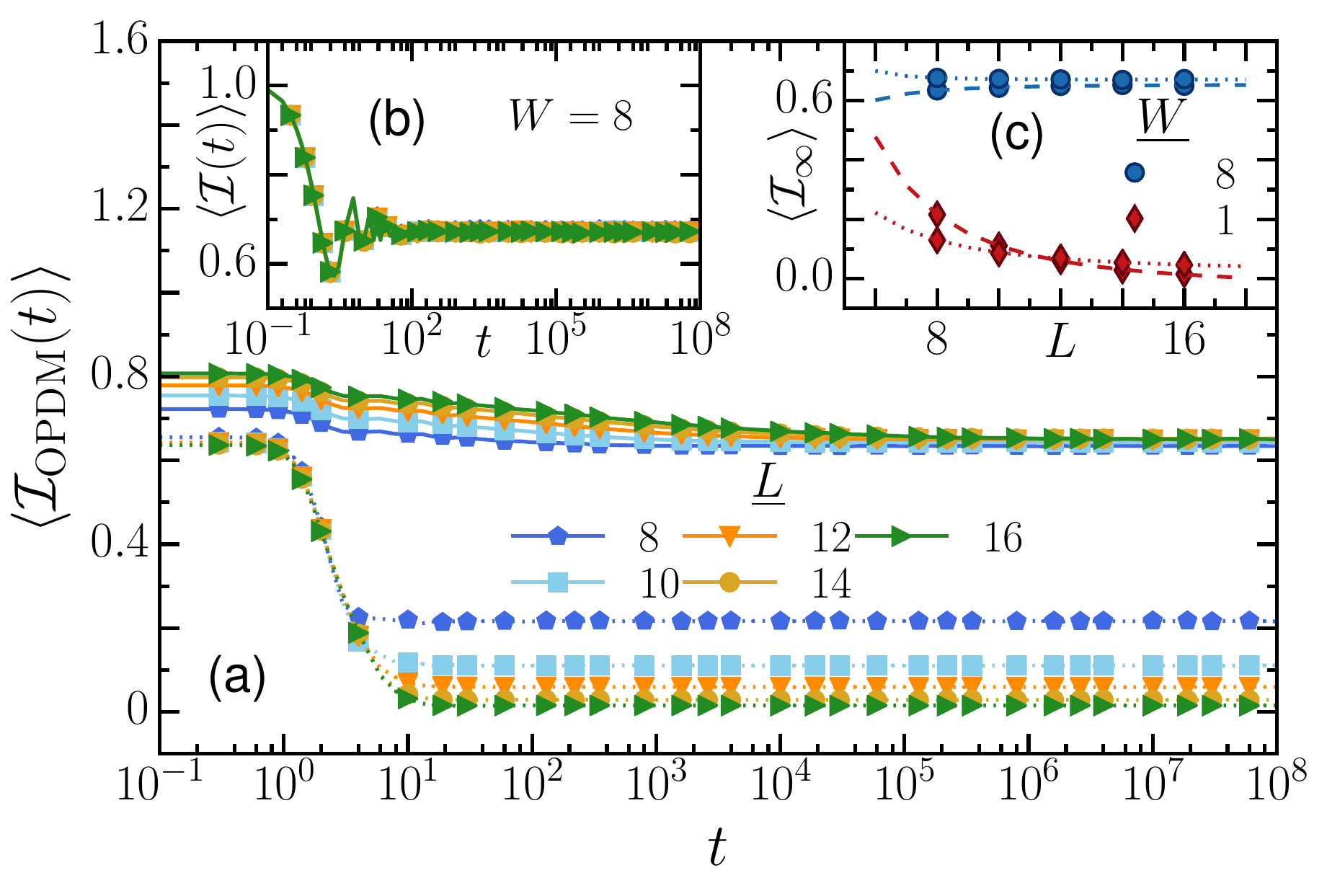}
\caption{
(a) Time-evolution of the disorder-averaged  OPDM occupation imbalance, $\langle\mathcal{I}_\mathrm{OPDM}(t)\rangle$, for $W=8$ (solid lines) and $W=1$ (dotted lines).
(b) Density imbalance $\langle\mathcal{I}(t)\rangle$ as a function of $L$ for $W=8$. 
(c) Finite-size scaling of the instantaneous disorder-averaged density imbalances at fixed large time $t\sim10^{8}$; $\langle\mathcal{I}_{\infty}\rangle$
($\langle\mathcal{I}(\infty)\rangle$: dotted lines;  $\langle\mathcal{I}_\mathrm{OPDM}(\infty)\rangle$: dashed lines. Both for $W=1,8$.)} 
\label{imbalance}
\end{figure}

{\it Imbalance and connection to experiments.}--The density imbalance $\mathcal{I}=(N_\text{e}-N_\text{o})/N$ between the number of physical particles $N_\text{e}$ on even sites and $N_\text{o}$ on odd sites is experimentally seen to relax to zero in the ergodic phase, whereas in the many-body localized phase it exhibits a fast relaxation towards a nonzero value, reflecting the absence of thermalization~\cite{Schreiber:2015jt}; 
similar conclusions were obtained numerically in Ref.~\onlinecite{Luitz2016}.
From the occupation spectrum, we define a related imbalance between the occupied and unoccupied halves of the spectrum as
\begin{equation}
\mathcal{I}_\text{OPDM}(t)=\frac{\langle N_{+}(t)\rangle-\langle N_{-}(t)\rangle}{N}, 
\label{imbl}
\end{equation} 
where $N_{+}(t) = \sum_{\alpha =1}^{N}\big(n_{\alpha}(t)-n_{\alpha}^{\mathrm{TH}}\big)$ and $N_{-}(t) = \sum_{\alpha =N+1}^{L}\big(n_{\alpha}(t)-n_{\alpha}^{\mathrm{TH}}\big)$.
We can view $\mathcal{I}_{\mathrm{OPDM}}(t)$ as a measure of how close a state is to a step-function occupation spectrum where the imbalance is maximal, or to a thermal occupation spectrum where its value is zero. 
This imbalance is plotted for both phases in \figref{imbalance}(a); it saturates to zero in the ergodic phase (dotted lines) and to a finite value in the MBL phase (solid lines). 
This conclusion is supported by the finite-size scaling given in \figref{imbalance}(c). 
In comparison with the density imbalance $\mathcal{I}(t)$ in the MBL phase, plotted in \figref{imbalance}(b), 
the relaxation of the OPDM imbalance is much slower.
The reason for this is that $\mathcal{I}(t)$ only captures the ballistic expansion part of the relaxation, while $\mathcal{I}_\mathrm{OPDM}(t)$ also captures the dephasing mechanism originating  from interactions between quasiparticles.
As a second main result of our work, we have thus demonstrated that the slow relaxation of the OPDM occupation spectrum can be directly connected to lack of ergodicity and experimentally accessible quantities.

{\it Conclusions.}---We demonstrated that in the many-body localized phase, the steady state reached at long times  
after a quench from a local product state has a smooth OPDM occupation spectrum with a highly nonthermal shape, in contrast to that obtained in the ergodic phase.
This is consistent with the picture of local conserved quantities which have a significant overlap with the initial state.
The approach towards the steady state is consequently a power law, reflecting dephasing via interactions between quasiparticles.
We have also defined an occupation imbalance, similar to the density imbalance used in experiments, that captures the main effect of dephasing and absence of thermalization.

Our discussion suggests that the continuous occupation spectrum is phenomenologically similar to that of a finite-temperature Fermi liquid.
The finite temperature is provided by the energy difference between the initial state and the closest eigenstate, which serve as reference states with a (zero-temperature) Fermi-liquid-like occupation spectrum.
This effective temperature can be tuned by changing the local structure of the initial state.
This is not thermalization in the conventional sense since the many-body localized phase is manifestly nonergodic.
Nevertheless, the observation that the steady state OPDM spectrum is continuous hints at the possibility of describing it with an 
emergent temperature, capturing the energy variance in the system.
It remains an interesting future research direction to establish whether such an emergent temperature corresponds to some thermal-like ensemble, necessarily different from eigenstate thermalization, and then how one can characterize it.
\\

\acknowledgements
This work was supported by the ERC Starting Grant No.~679722 and EPSRC grant No. EP/P010180/1.
The work of F.H.-M. was performed in part  
at the Aspen Center for Physics 
which is supported by National Science Foundation
Grant No. PHYS-1607611. The hospitality of the Aspen Center for Physics is gratefully acknowledged.

\bibliography{references} 

\appendix

\section{Time average of the OPDM and smearing of its occupation spectrum in the MBL phase.}

\begin{center}
\begin{figure*}[t]
\centering
\includegraphics[width=0.32\textwidth]{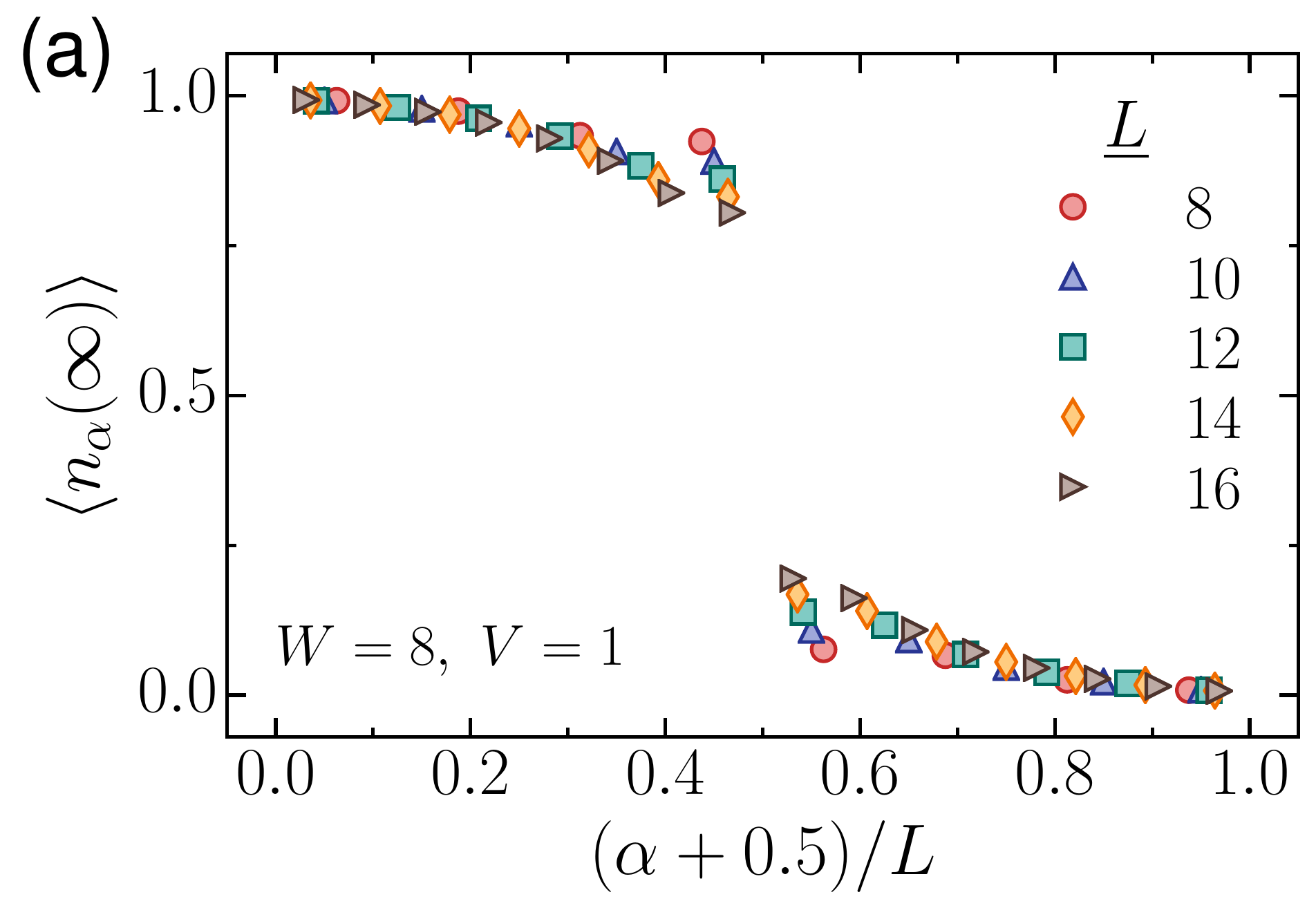}
\includegraphics[width=0.32\textwidth]{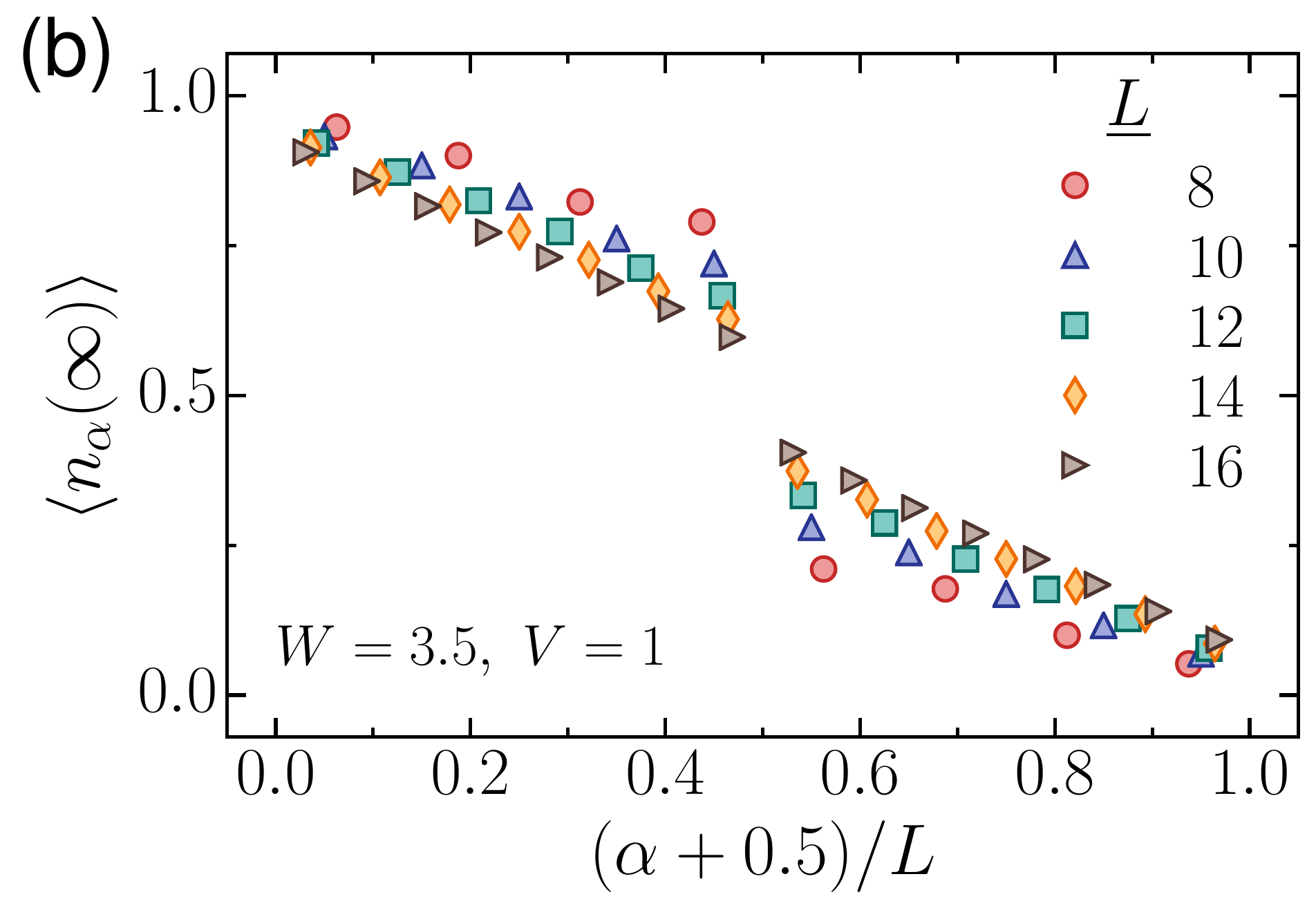}
\includegraphics[width=0.32\textwidth]{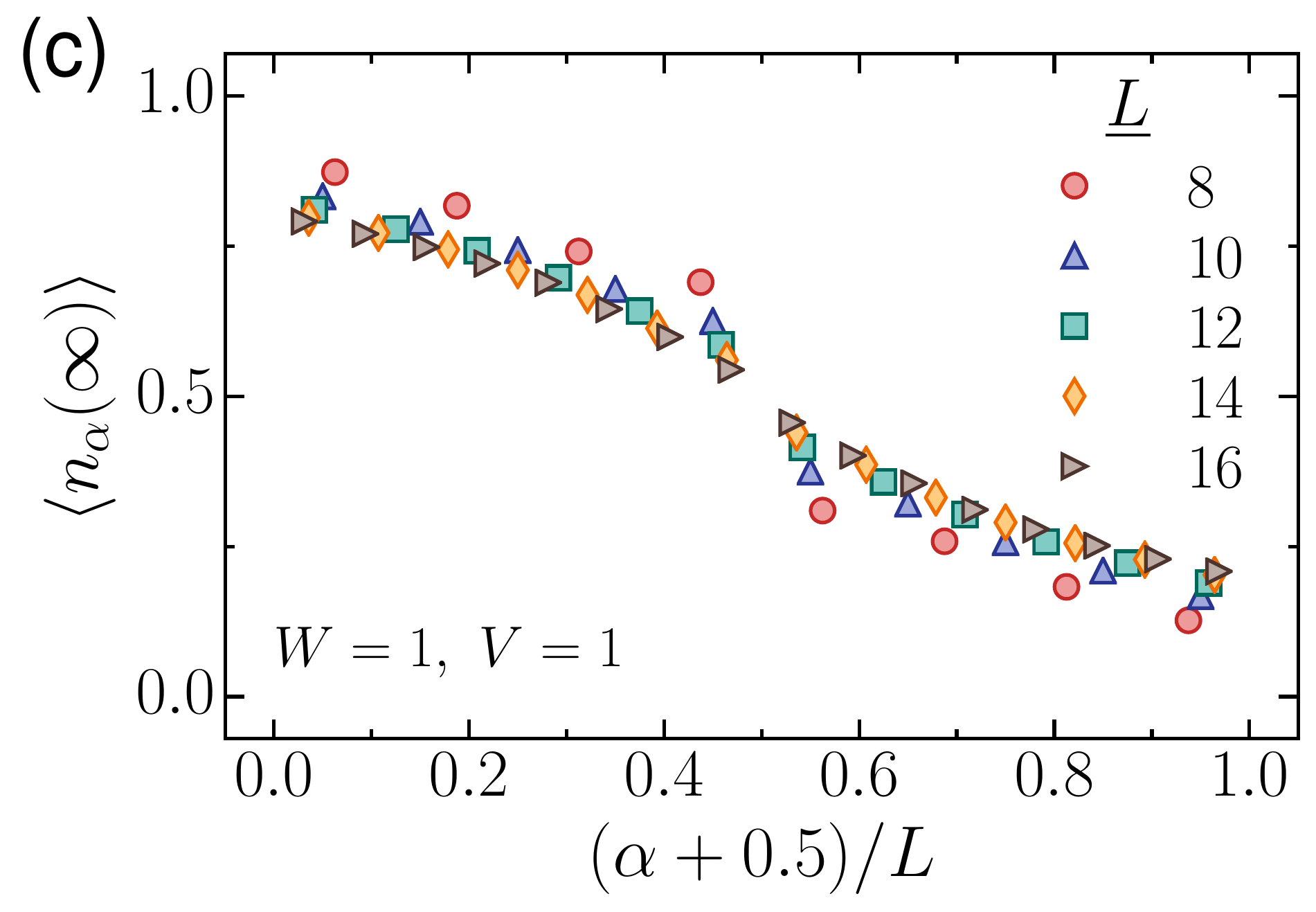}
\caption{Instantaneous disorder-averaged distribution of occupations $\langle n_{\alpha}(\infty)\rangle$ at fixed large time $t=10^8$, as a function of system size $L$ and disorder strengths: (a) $W=8$, (b) $W=3.5$ and (c) $W=1$.  The horizontal axis is scaled to $(\alpha+0.5)/L$.
}
\label{na}
\end{figure*}
\end{center}

In this appendix we provide more details on some technical points mentioned in the main text, starting with the time average of the density matrix. 
First, in \figref{na} we plot the  instantaneous OPDM spectrum at large times ($t = 10^{8}$), obtained by first ordering the occupations by size and then taking the disorder average.
This should be compared with the time-averaged OPDM spectrum in Fig.~2 of the main text.
We notice that the instantaneous occupations are systematically larger, although as the system size becomes larger, the distribution in the thermal phase approaches its infinite-time average, whereas the distribution in the MBL phase does not.
As already mentioned in the main text, this has mainly to do with the fact that the ordering of eigenvalues does not commute with the disorder average when occupations are interchanged as a function of time.

\begin{center}
\begin{figure}[t]
\includegraphics[width=0.4\textwidth]{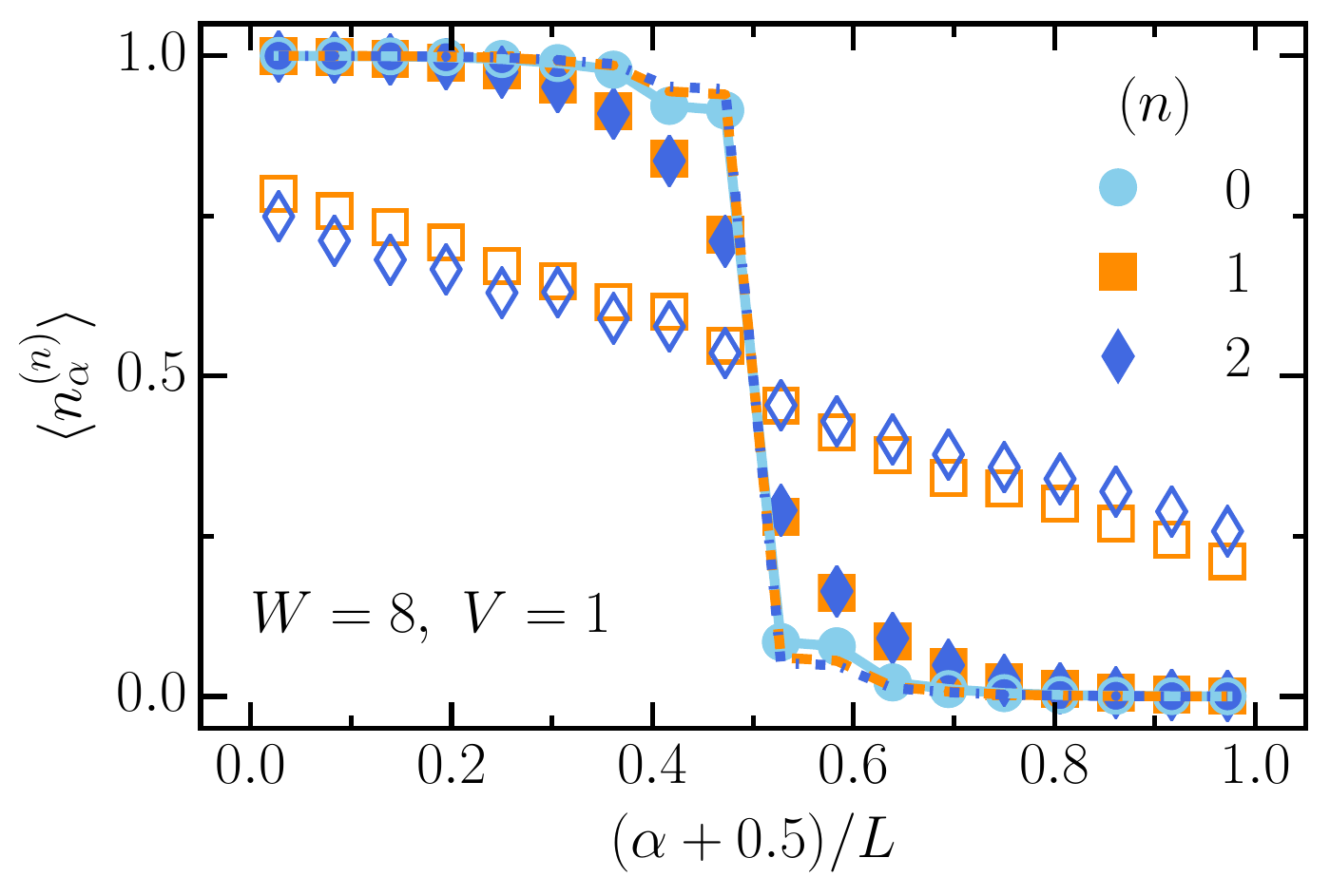}
\caption{Disorder-averaged occupations $\langle n^{(n)}_{\alpha}\rangle$ in the three many-body eigenstates ($n=0,1,2$) with the largest weights in Eq.~(7). The horizontal axis is scaled to $(\alpha+0.5)/L$ and $L=18$. Filled and unfilled symbols are the $n^{(n)}_{\alpha}$ ordered (in descending order)  {\it before} and {\it after} taking the disorder average, respectively (which yields the same result for $n=0$.) The solid, dashed and dashed-dotted lines are the disorder-averaged occupations in the eigenstates $n=0,1,2$, respectively.
}
\label{fig:naprox}
\end{figure}
\end{center}

Second, we give more details on the arguments leading to Eq.~(8) in the main text.
The maximum-overlap eigenstate $\ket{0}$ has an OPDM $\rho^{(0)}$ with a zero-temperature Fermi-liquid-like step function, as we confirm in \figref{fig:naprox}, where we plot its eigenvalues $n_\alpha^{(0)}$ (the same is the case for the other eigenstates, see lines).
The unitary transformation $U_{0}$ that diagonalizes $\rho^{(0)}$ approximately diagonalizes the OPDM $\rho^{(n)}$ of the higher eigenstates $\ket{n}$, such that the diagonal elements $n_\alpha^{(n)} = (U_{0}^{-1}\rho^{(n)}U_{0})_{\alpha\alpha}$, if ordered in decreasing order before taking the disorder average, also have an OPDM occupation discontinuity (see filled symbols in \figref{fig:naprox}).
The transformation $U_{0}$, however, randomizes the ordering. 
To demonstrate this, the occupations in the zeroth, first and second state $\{n_{\alpha}^{(n)}:n=0,1,2\}$ are plotted for three disorder configurations in \figref{napsr}. 
As a consequence, the disorder average $\langle n_\alpha^{(n)} \rangle$, plotted with unfilled symbols in \figref{fig:naprox}, becomes a smooth function without any discontinuity.

Assuming then that the diagonal elements of the diagonal density matrix are obtained by applying $U_0$ term by term in Eq.~(6), we write   
\begin{equation}
n_{\alpha}^{\mathrm{diag}} = |a_{0}|^{2}n_{\alpha}^{(0)} + \sum_{n>0}|a_{n}|^{2}n_{\alpha}^{(n)}.
\label{eq:na_pert}
\tag{S1}
\end{equation}
Here, as in the main text, the $\{\ket{n}\}$ are ordered in order of decreasing $|a_n|$.
Given that all the higher order terms in Eq.~\eqref{eq:na_pert} are smooth, even at finite system sizes, they do not contribute to the discontinuity $\Delta n$, which is therefore given by Eq.~(8) in the main text.
In \figref{wsapl} we compare the discontinuity $\Delta n$ obtained from the time average with that obtained from the approximation of Eq.~(8).
The agreement between the two quantities is rather good, given the approximations used in the argument leading up to the prediction of Eq.~(8), and corroborates the conclusion that in the thermodynamic limit, the discontinuity indeed goes to zero.

\section{Reference initial states and their overlap with the many-body eigenstates.}
In the main text, we stated the importance of selecting initial states that resemble zero-temperature reference states (many-body eigenstates). 
We also provided a way to produce such a class of initial states $\ket{\Psi_{\gamma}}$, and compared them with an initial state $\ket{\Psi_{\mathrm{free}}}$ that is not representative of such reference states. Here, we show in \figref{overlap}, how the local structure of the initial states is reflected in their overlap $\langle|a_n|^{2}\rangle$ with the many-body eigenstates $\ket{n}$ (in line with the analysis in Fig.~2(c) in the main text). As we increase $\gamma$, the largest overlap $\langle|a_0|^{2}\rangle$ between the initial states $\ket{\Psi_{\gamma}}$ and the eigenstates decreases, but even for large $\gamma$, there is an eigenstate with a significant weight that serves as a reference. In contrast, the overlaps of $\ket{\Psi_{\mathrm{free}}}$ with $\ket{n}$ are essentially constant.
\begin{onecolumngrid}
\begin{center}
\begin{figure}[h]
\includegraphics[width=0.32\textwidth]{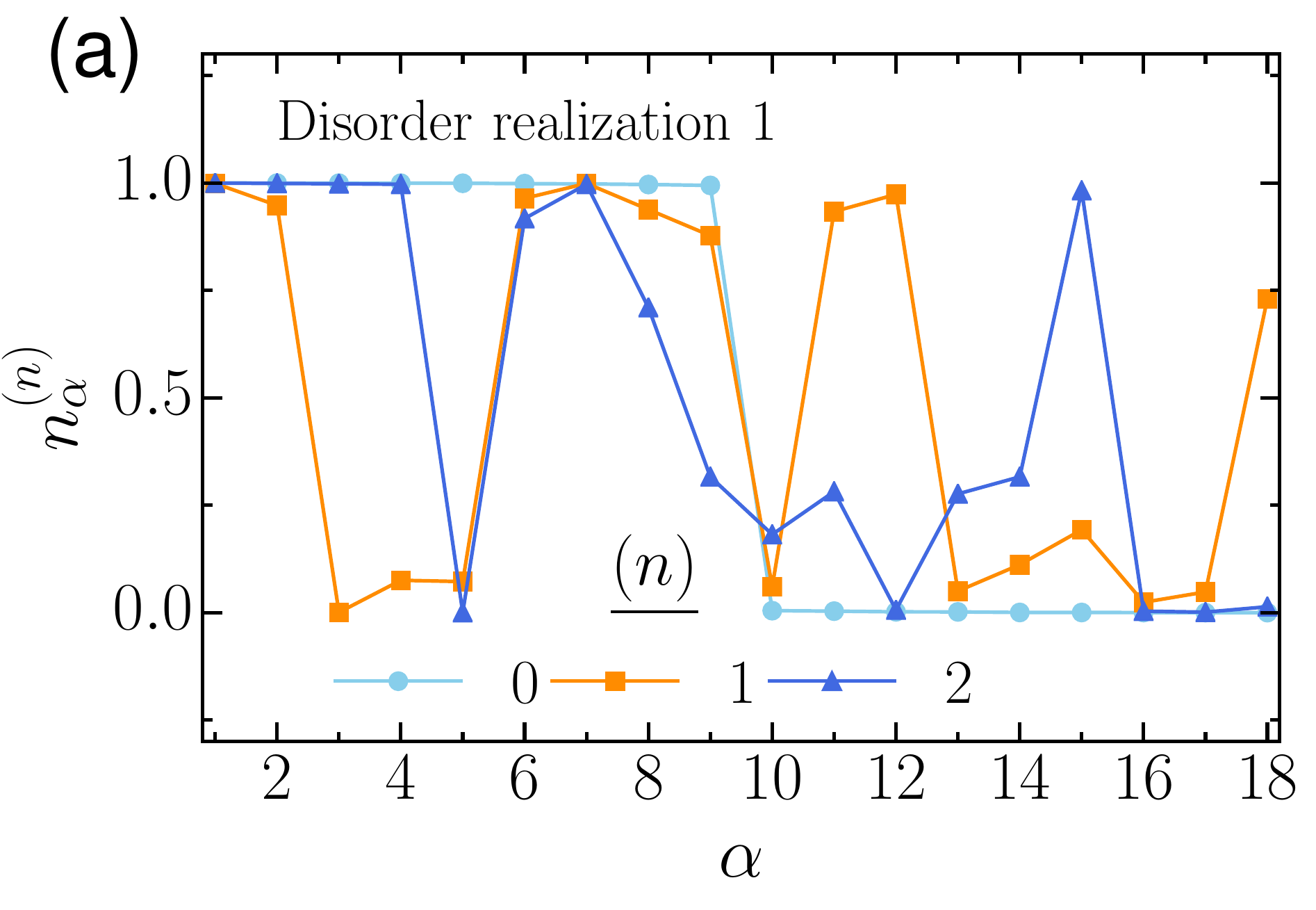}
\includegraphics[width=0.32\textwidth]{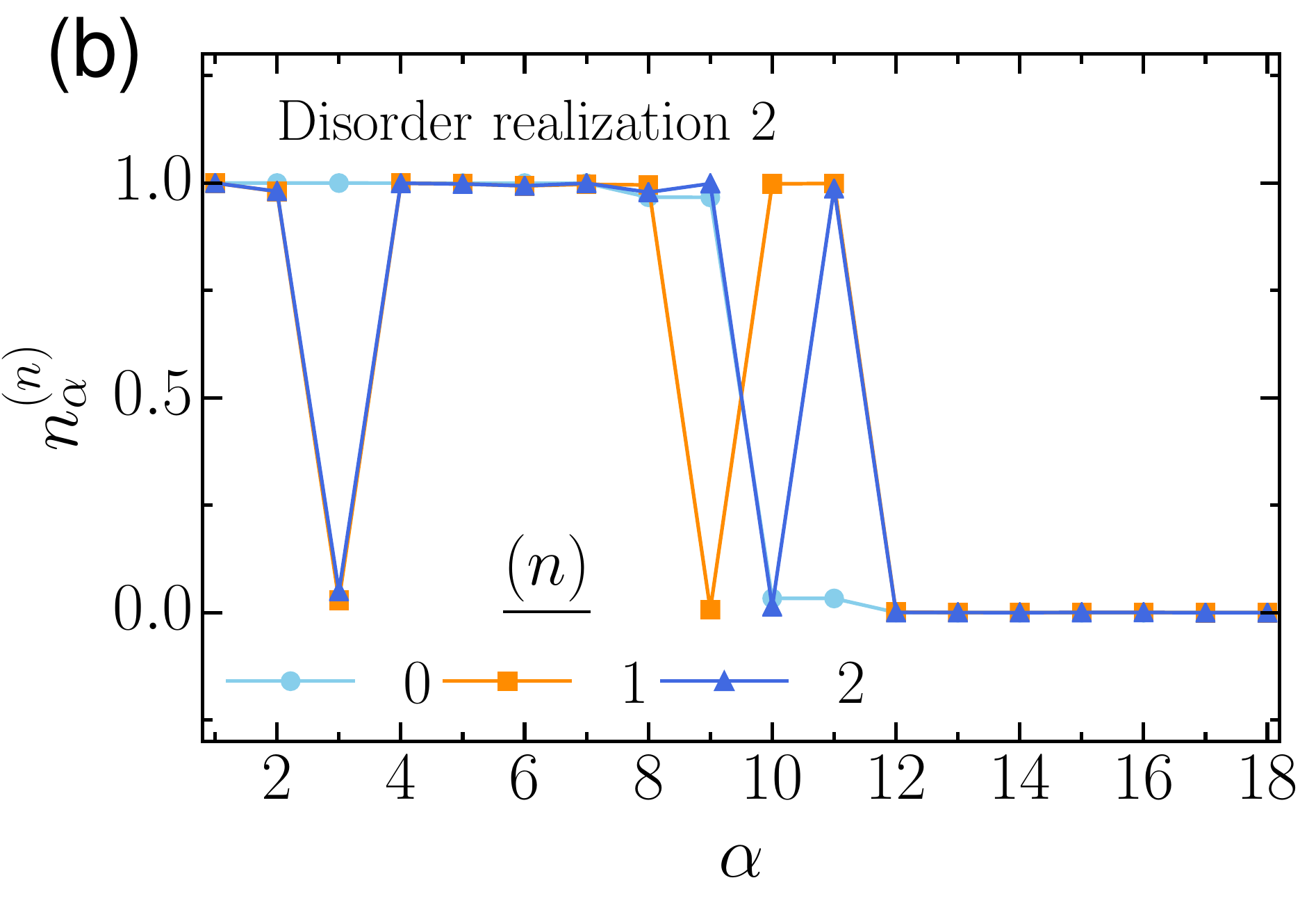}
\includegraphics[width=0.32\textwidth]{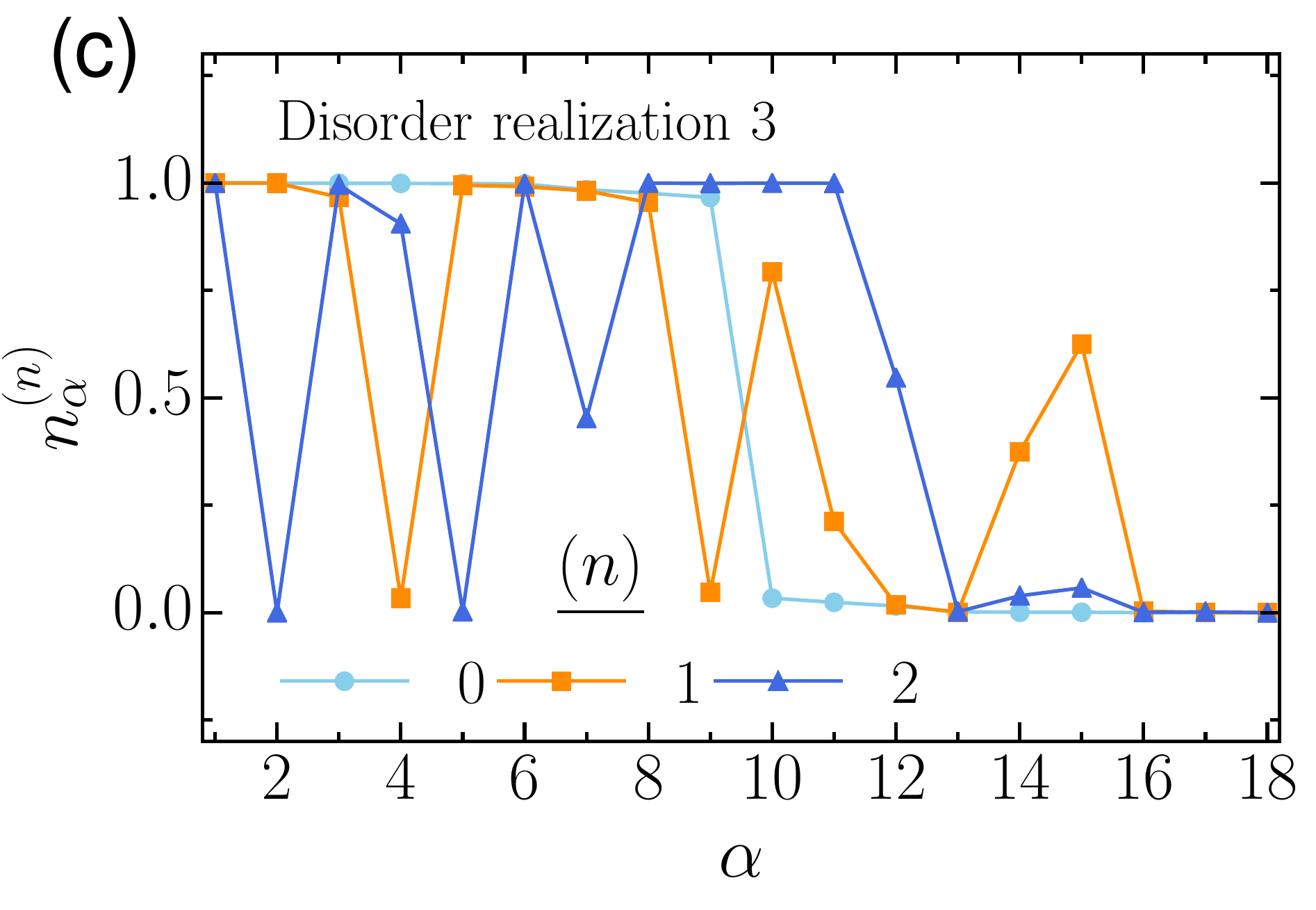}
\caption{Diagonal elements $\{n_{\alpha}^{(n)}:n=0,1,2\}$ of the three many-body eigenstates with the largest weights $|a_n|$ in Eq.~\eqref{eq:na_pert} for $W=8$ as a function of $\alpha$, obtained after a unitary transformation with the matrix $U_0$ that diagonalizes the largest-weight eigenstate $\ket{0}$.  Data is shown for three different random disorder configurations in (a), (b) and (c).}
\label{napsr}
\end{figure}%
\begin{figure}[H]
    \begin{minipage}{0.48\textwidth}
     \centering
        \includegraphics[width=0.8\textwidth]{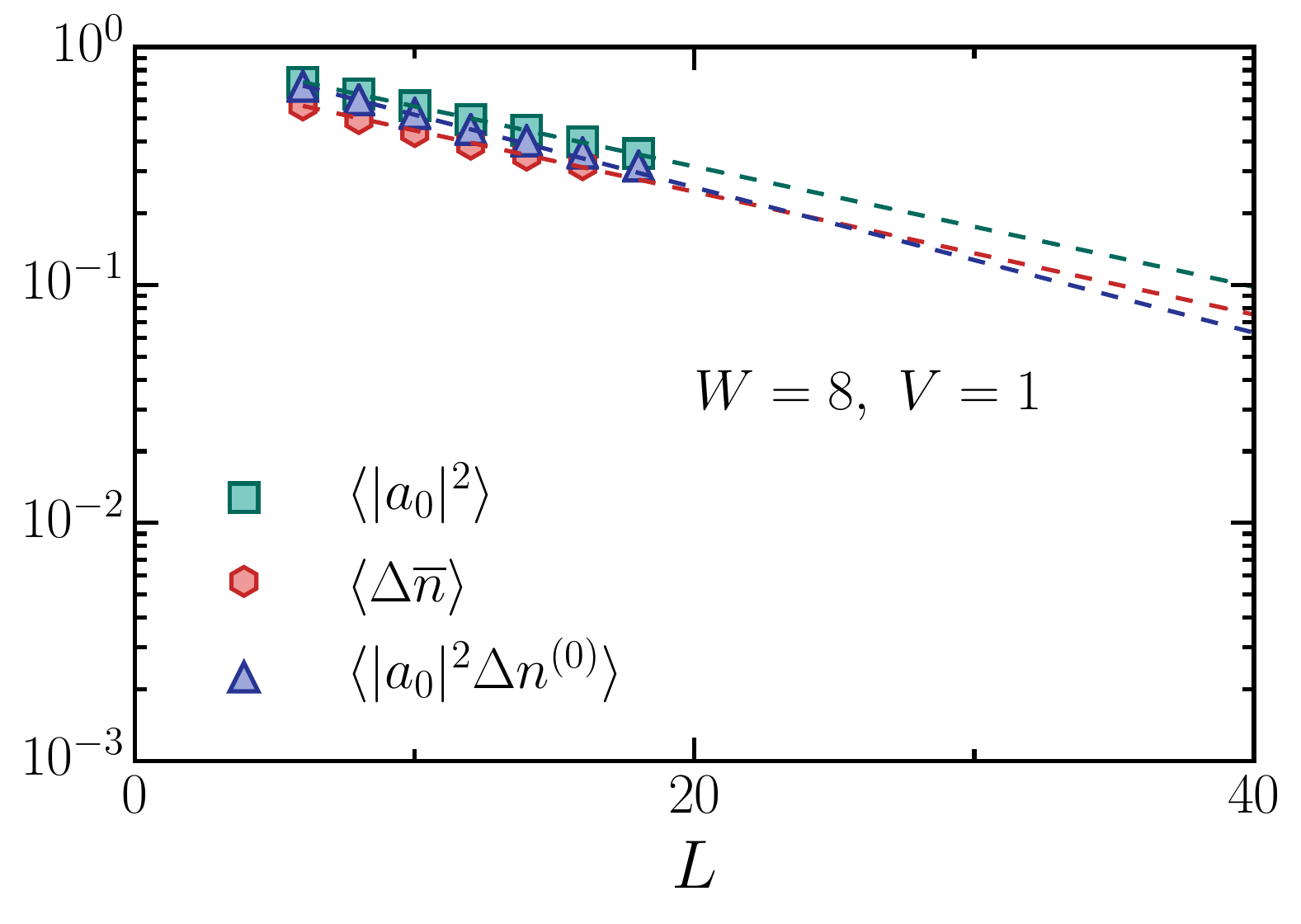}
       \caption{Disorder-averaged largest eigenstate weight $\langle|a_{0}|^{2}\rangle$, infinite-time and disorder-averaged discontinuity $\langle\Delta\overline{n}\rangle$ and its zeroth-order approximation $\langle|a_{0}|^{2}\Delta n^{(0)}\rangle$, for $W=8$ as a function of system size $L$ and a fit to an exponential  $\sim e^{-b L}$, with $b\approx 0.06$ for data in squares and triangles, and $b\approx0.07$ for data in octagons. }
\label{wsapl}
    \end{minipage}%
    \hfill%
    \begin{minipage}{0.48\textwidth}
    \centering
        \includegraphics[width=0.8\textwidth]{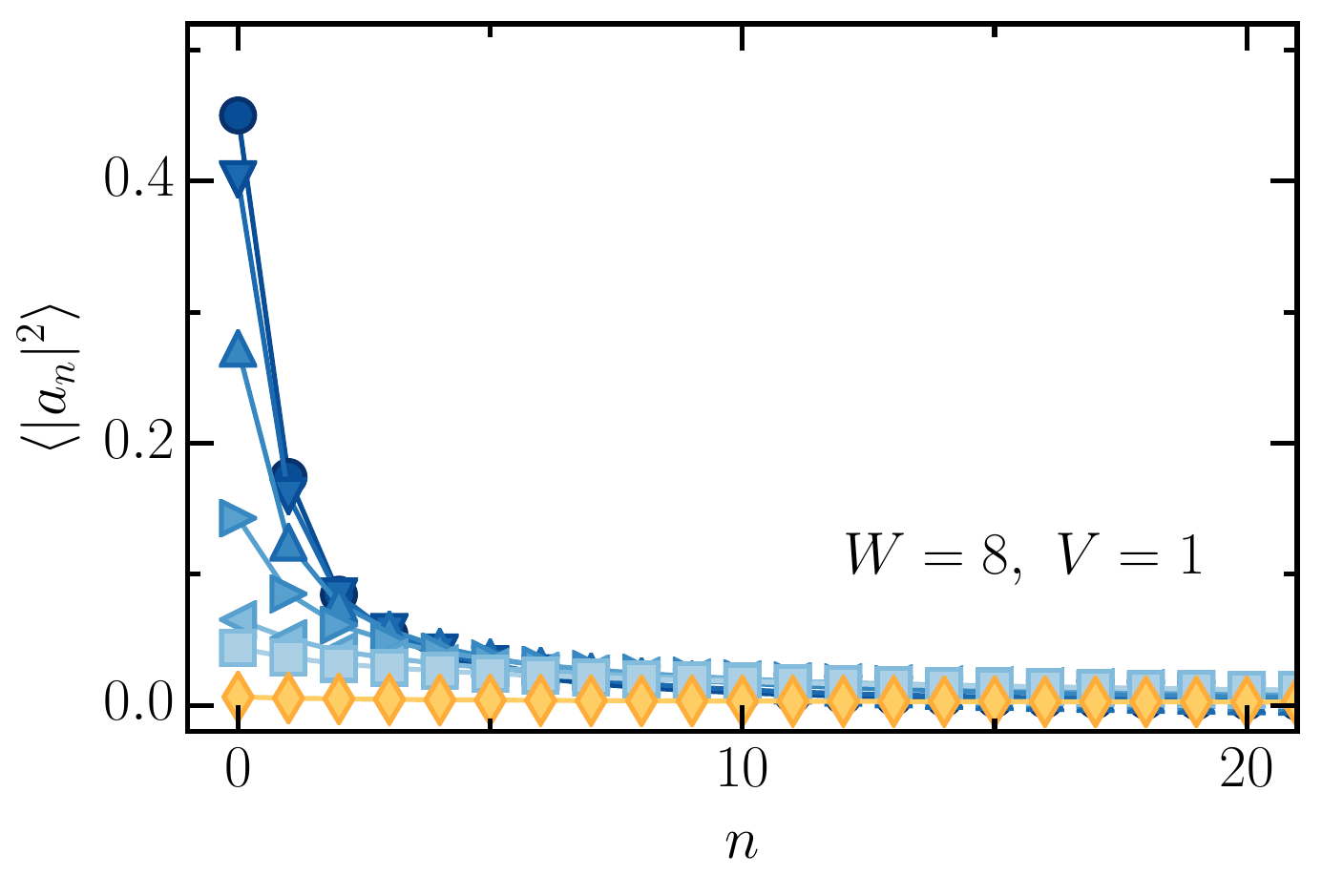}
       \caption{Disorder-averaged overlap $\langle |a_{n}|^{2} \rangle$ between the initial states of the form $|\Psi_{\gamma}\rangle$ with $\gamma= m\pi/20$; $m=0,1,\dots,5$ (from top to bottom) and the many-body eigenstates $\ket{n}$, plotted in decreasing order $|a_0|>|a_1|>\dots$ as a function of $n$, for $W=8$ and $L=14$. The same is plotted for the clean initial state $|\Psi_{\mathrm{free}}\rangle$ in diamonds.}
\label{overlap}
    \end{minipage}
 \end{figure}
\end{center}%
\end{onecolumngrid}

\end{document}